\documentclass[twocolumn]{aastex631}

\usepackage[normalem]{ulem}
\usepackage{color} 

\def\sagii{Sagittarius~II}
\def\ufd{UFD}

\received{May 17, 2024}
\shorttitle{UFDs challenge the CDM paradigm}
\shortauthors{S\'anchez Almeida et al.}

\begin{document}

\title{
The stellar distribution in ultra-faint dwarf galaxies suggests deviations from the collision-less cold dark matter paradigm} 

\correspondingauthor{JSA}
\email{jos@iac.es}

\author[0000-0003-1123-6003]{Jorge S\'anchez Almeida} \affil{Instituto de Astrof\'\i sica de Canarias, La Laguna, Tenerife, E-38200, Spain} \affil{Departamento de Astrof\'\i sica, Universidad de La Laguna}

\author[0000-0001-8647-2874]{Ignacio Trujillo} \affil{Instituto de Astrof\'\i sica de Canarias, La Laguna, Tenerife, E-38200, Spain} \affil{Departamento de Astrof\'\i sica, Universidad de La Laguna}

\author[0000-0001-5848-0770]{Angel R. Plastino} \affil{CeBio y Departamento de Ciencias B\'asicas, \\ Universidad Nacional del Noroeste de la Prov. de Buenos Aires, \\ UNNOBA, CONICET, Roque Saenz Pe\~na 456, Junin, Argentina}







\begin{abstract}
Unraveling the nature of dark matter (DM) stands as a primary objective in modern physics. Here we present evidence suggesting deviations from the collisionless Cold DM (CDM) paradigm.  It arises from the radial distribution of stars in six Ultra Faint Dwarf (UFD) galaxies  measured with the Hubble Space Telescope (HST). After a trivial renormalization in size and central density, the six UFDs show the same stellar distribution, which happens to have a  central plateau or {\em core}. Assuming spherical symmetry and isotropic velocities, the Eddington inversion method proves the observed distribution to be inconsistent with potentials characteristic of CDM particles. Under such assumptions, the observed innermost slope of the stellar profile discards the UFDs to reside in a CDM potential at a $\geq 97$\,\%\ confidence level.  The extremely low stellar mass of these galaxies, $10^3$\,--\,$10^4~{M_\odot}$, prevents stellar feedback from modifying the shape of a CDM potential. Other conceivable explanations for the observed cores, like deviations from spherical symmetry and isotropy, tidal forces, and the exact form of the used CDM potential, are disfavored by simulations and/or observations. Thus, the evidence suggests that collisions among DM particles or other alternatives to CDM are likely shaping these galaxies. Many of these alternatives produce cored gravitational potentials, shown here to be consistent with the observed stellar distribution. 
\end{abstract}



\section{Introduction} \label{sec:intro}

The shape of the DM haloes of low-enough mass galaxies encodes direct information on the nature of DM. Self gravitating collisionless Cold DM (CDM) halos evolving in a cosmological context develop a central cusp where the mass density profile increases toward the center
following the NFW profile or other similar shape (after Navarro, Frenk, and White \citeyear{1997ApJ...490..493N}; see also \citeauthor{2020Natur.585...39W}~\citeyear{2020Natur.585...39W}). When baryons are included, baryonic processes can thermalize the overall gravitational potential turning the central cusp into a plateau or core \citep[e.g.,][]{2010Natur.463..203G}, a mechanism invoked to explain the DM haloes observed in dwarf galaxies \citep[e.g.,][]{2015AJ....149..180O}.  The energy needed to turn cusps into cores must be extracted from the star-formation, therefore, when  the formed stellar mass is too small,  the baryon feedback is unable to transform cusps into cores and the DM haloes remain cuspy. The actual largest stellar mass ($M_\star $) unable to modify the CDM potential is model dependent \citep[][]{2016MNRAS.459.2573R} but it roughly corresponds to $M_\star < 10^{6}\, M_{\odot}$ or to a DM halo mass $< 10^{10}\,M_\odot$  \cite[][]{2014MNRAS.437..415D,2015MNRAS.454.2981C,2020ApJ...904...45H,2021MNRAS.502.4262J}. 
Thus, if the DM haloes of these Halo Unevolved Galaxies (HUGs) happen to show a core, it would indicate the  DM not being collisionless, reflecting the much sought-after presently-unknown true nature of the DM \citep[fuzzy, self-interacting, warm, or other alternatives; e.g.,][]{2022arXiv220307354B}.

In practice, DM halo shapes are deduced from spatially-resolved kinematical measurements, which require time-consuming high spectral resolution spectroscopy and are  virtually imposible in the required HUG regime. \citet[][]{2023ApJ...954..153S} proposed an alternative based on photometry, much cheaper observationally, but starting from a series of simplifying assumptions that must be justified a posteriori.  It uses the classical Eddington Inversion Method \citep[EIM; e.g., ][]{2008gady.book.....B,2021isd..book.....C}, which provides the distribution function (DF) to be followed by a  mass density profile immerse in an spherically symmetric gravitational potential. If the required DF becomes negative somewhere in the phase space, it proves the pair density--potential to be physically inconsistent with each other. Such inconsistency happens for a combination particularly interesting in the present context, namely, a stellar density with a core residing in a NFW potential \citep[][]{2006ApJ...642..752A,2009ApJ...701.1500A,2023ApJ...954..153S}.  Stellar cores seem to be quite common in low mass galaxies \citep[e.g.,][]{2020ApJ...892...27M,2021ApJ...922..267C} and, if their presence remains in the critical HUG mass range (provided they meet the requirements of EIM), it could indicate the need to go beyond the standard CDM model.

Here we analyze the stellar count distribution of six UFD satellites of the Milky Way (MW) and the Large Magellanic Cloud (LMC) from \citet{2024ApJ...967...72R}. Their stellar masses are  in the interesting HUG regime, $10^3$\,--\,$10^4\, M_\odot$, and they present stellar surface density profiles with cores. Here we consider whether these facts represent evidence for the DM deviating from the CDM paradigm. The work is presented as follows: Sect.~\ref{sec:observations} summarizes the observations and shows that the same radial profile reproduces all galaxies simultaneously within the error set by star counting. Section~\ref{sec:derivation} outlines the EIM-based procedure used to infer the DF needed to explain the observed profile assumed a potential. In Sect.~\ref{sec:results}, the procedure is applied to conclude that NFW potentials require unphysical negative DF and provide fits significantly worst than potentials with cores (Schuster-Plummer and $\rho_{230}$ potentials). 
The conclusion that the satellites do not reside in NFW potentials depends on several simplifications and assumptions: steady state, stacking of profiles, spherical symmetry, shape of the potential, isotropic velocities,  and unimportance of tidal forces and stellar feedback effects. All of them are discussed in Sect.~\ref{sec:conclusions}, and the most compelling explanation for the existence of stellar cores in these dwarf galaxies remains a deviation from the CDM paradigm.

%
\section{Data}\label{sec:observations}
%
\citet{2024ApJ...967...72R} studied ten UFD satellites of the MW and the LMC. Deep  HST two-band photometry  (F606W and F814W) allows them to select stars individually and to separate them from foreground and background contaminants. Only six of the UFDs have  a field-of-view large enough to have good determination of structural parameters, and these are the targets employed in the present study: Horologium~I, Horologium~II, Hydra~II, Phoenix~II, Sagittarius~II,  and Triangulum~II. The observed stellar counts of each galaxy were modeled using 2D Schuster-Plummer functions and exponentials, leaving free parameters that include the center, the characteristic radius, the ellipticity, and the orientation. These 2D fits were later used to construct 1D radial profiles as the number density of counts in ellipses with the ellipticity, orientation, and center of the best fitting 2D functions. We use these 1D profiles in our work, both from the Schuster-Plummer function and the exponential since their differences quantify the systematic errors induced by the determination of centers, ellipticities, and orientations. The error in each radial bin is estimated as the Poisson noise arising from star counts.

From the above fits and ancillary data,  \citet{2024ApJ...967...72R} show the \ufd s to have axial ratios from 0.55 to 1, $M_\star$ from $6\times 10^2$ to $2.4\times 10^4\,M_\odot$, and dynamical masses at the half-light radius between $10^5$ and $5\times 10^6\,M_\odot$. Thus, the ratio of dynamical mass to stellar mass within the half-light radius goes from 300 to 3000, with the exception of  \sagii\ where it is {\em only} around 10.  It has been argued that \sagii\  may be a globular cluster (GC) of large size \citep{2021MNRAS.503.2754L} but from the point of view of its surface density profile, it behaves as the rest. It is analyzed together with the others and separately, as discussed in Sect.~\ref{sec:conclusions}.
For further details on the dataset and reduction, we refer to \citet{2024ApJ...967...72R}.

\begin{figure*}[ht!] 
\centering
\includegraphics[width=0.8\linewidth]{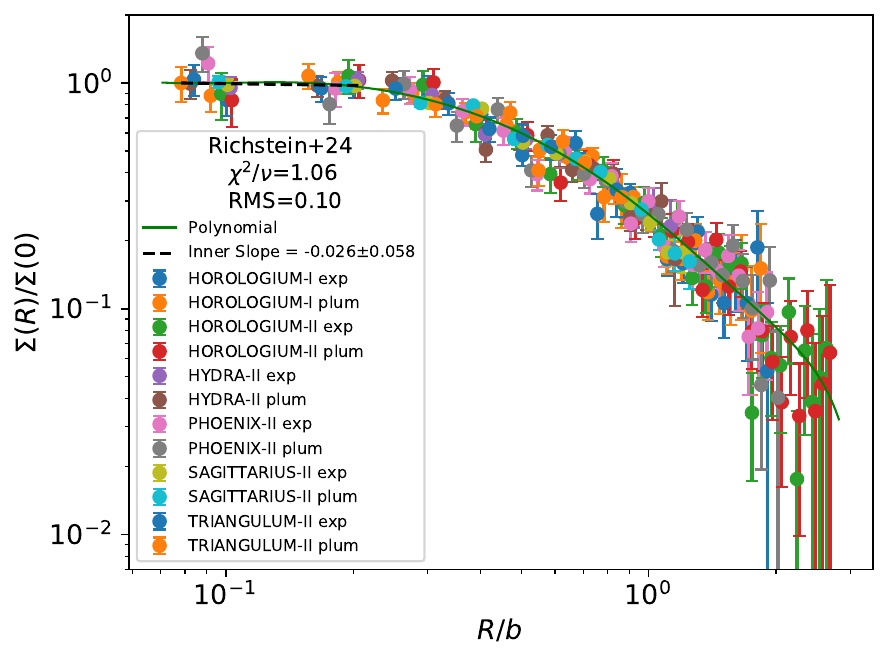}
\caption{
Stellar surface density profiles for six \ufd s from \citet{2024ApJ...967...72R}, pieced together by rescaling in $x$ and $y$ using a least-square fit  that gives the scales and assumes the same polynomial shape for all galaxies. The symbols represent  different galaxies and reductions,  as labeled, whereas the solid line shows the best fitting polynomial.  The dashed line corresponds to a linear fit to the innermost points of the profile.  The best fitting polynomial is close to a simple $y=1/(1+x^2)$ law (not shown). $\chi^2/\nu$ stands for the reduced $\chi^2$ of the fit with RMS representing its root mean square.
}
\label{fig:read_richstein5}
\end{figure*}

Figure~\ref{fig:read_richstein5} shows the 1D stellar surface density of the six UFDs normalized in $x$ (size) and $y$ (central density) using a least squares procedure to set the scales so that the resulting profile, assumed to be a polynomial, is the same for all\footnote{$\log\left[\Sigma(R)/\Sigma(0)\right]=\sum_{i=0}^5 c_i \left[\log(R/b)\right]^i$, with $c_0, \dots, c_5=$ 0.262, $-0.907$,  1.019,  0.640 , $-1.619$,  $-1.072$, valid for $0.07 \leq R/b \leq 2.7$. The symbols $b$ and $\Sigma(0)$ stand for the scales in $x$ and $y$, respectively. The half-mass radius is $0.52b$ whereas the core radius, defined as in Eq.~(\ref{eq:tricks}), is $0.60b$.}. The reduced $\chi^2/\nu$ of the fit\footnote{$\chi^2$ is the sum of the error normalized squares of the residuals whereas $\nu$ represents the degrees of freedom.}, $1.06$, implies that all galaxies are well reproduced with a single profile with the scatter almost exclusively set by the star counting. This result is remarkable and implies that the profile in Fig.~\ref{fig:read_richstein5} represents galaxies with any ellipticity including completely round ones as expected in spherically symmetric systems in steady state. Moreover, this common profile has a central plateau or {\em core} whose logarithmic slope when $R\to 0$ is,
\begin{equation}
\omega \equiv   \frac{d\log \Sigma(R)}{d\log R} \simeq -0.026\pm 0.058,
  \label{eq:innermostslope}
\end{equation}
determined from a linear fit to the 12 innermost points, chosen because they delineate the plateau (the black dashed line in Fig.~\ref{fig:read_richstein5}). The normalized data points in Fig.~\ref{fig:read_richstein5} are the observation analyzed in this work. It represents stellar counts but we use them as proxies for the stellar mass density distribution, which is a good approximation for the UFDs having old stellar populations \citep[e.g.,][]{2021ApJ...920L..19S}.

%
\section{Eddington inversion method approach}\label{sec:derivation}

The details and tests of the technique are given elsewhere \citep{sanchez_almeida_nube}, but here we summarize the approach used to compute the DF in the phase-space $f$ required for the observed profile (Fig.~\ref{fig:read_richstein5}) to reside in a particular potential. For a spherically symmetric system of identical stars with isotropic velocity distribution, $f(\epsilon)$ depends only on the particle energy $\epsilon$. (The impact of relaxing these assumptions is addressed in Sect.~\ref{sec:conclusions}.) Then, the stellar volume density $\rho(r)$ turns out to be \citep[e.g.,][Sect.~4.3]{2008gady.book.....B},
\begin{equation}
  \rho(r) = 4 \pi \sqrt{2} \,\int_0^{\Psi(r)} \, f(\epsilon) \sqrt{\Psi(r) - \epsilon} \, d\epsilon,
  \label{eq:leading}
\end{equation}
with $\epsilon = \Psi - \frac{1}{2} v^2$ the relative energy per unit mass of a star and $\Psi(r) = \Phi_0 - \Phi(r)$ its relative potential energy.  The symbol $\Phi(r)$ stands for the gravitational potential energy  and $\Phi_0$ is $\Phi(r)$ evaluated at the edge of the system. The previous equation can be rewritten as
\begin{equation}
  \rho(r) = \int_0^{\epsilon_{max}}f(\epsilon)\,\xi(\epsilon,r)\, d\epsilon,
  \label{eq:master}
\end{equation}
with 
\begin{equation}
  \xi(\epsilon,r)=4 \pi \sqrt{2\epsilon_{\rm max}}\sqrt{\left[\frac{\Psi(r)}{\Psi(0)}-\frac{\epsilon}{\epsilon_{max}}\right]} \,\,\Pi(X-r),
  \label{eq:master_mind}
\end{equation}
$\epsilon_{max}= \Psi(0)$, $X$ the radius implicitly defined as $\Psi(X)/\Psi(0) = \epsilon/\epsilon_{max}$, and $\Pi(x)$ the step function,
\def\cacab{$x\leq 0$}
\def\cacac{$x > 0$}
\begin{equation}
  \Pi(x) =
  \cases{
    0 & \cacab ,\cr
    1 & \cacac.
    }
  \end{equation}
The symbol $\xi(\epsilon,r)$ represents a family of densities that are characteristic of the potential and dependent on the energy $\epsilon$. Then, according to Eq.~(\ref{eq:master}), the stellar density is just the superposition of these characteristic densities with the DF $f(\epsilon)$ giving the contribution of each energy to $\rho(r)$. (The characteristic densities for a Schuster-Plummer potential are shown as an example in Appendix~\ref{sec:plummer_pot}.) 
Following Eq.~(\ref{eq:master}), $f(\epsilon_i)$ could be retrieved by fitting the observable $\rho(r)$ with a linear superposition of $\xi(\epsilon_i,r)$ at various $\epsilon_i$. (We will see below that 
 $\rho$ can be replaced with the projected stellar surface density, which is the true observable.) In practice, however, there is no error-proof way to discretize Eq.~(\ref{eq:master}). We approach the practical problem by expanding $f(\epsilon)$ as a polynomial of order $n$,     
\begin{equation}
  f(\epsilon)\simeq \epsilon_{max}^{-3/2} \sum_{i=3}^n\,a_{i}\,(\epsilon/\epsilon_{max})^i,
  \label{eq:polydef}
\end{equation}
so that
\begin{equation}
  \rho(r) \simeq \sum_{i=3}^n\,a_i\,F_i(r),
  \label{eq:master2}
\end{equation}
\begin{equation}
  F_i(r) =\epsilon_{max}^{-1/2}\int_0^{1}\,\alpha^i\,\xi(\alpha\,\epsilon_{max},r)\,d\alpha,
  \label{eq:master3}
\end{equation}
with $\alpha = \epsilon/\epsilon_{max}$. Equation~(\ref{eq:master2}) gives a simple expansion of the stellar density $\rho(r)$ in terms of potential-dependent but known functions $F_i(r)$.
The chosen functional form in Eq.~(\ref{eq:polydef}) is both flexible and, by starting at $i=3$, it describes a system of finite mass despite
the mass given by $\xi(\epsilon,r)$ diverges as $\epsilon^{-\gamma}$ when $\epsilon\to 0$,
with  $2 < \gamma < 3$ depending on the potential \citep{sanchez_almeida_nube}.  
The normalization   in Eq.~(\ref{eq:polydef})  has been chosen so that $F_i(r)$ does not depend on $\epsilon_{max}$. The discretization in Eq.~(\ref{eq:master2}) also holds for the projection of the volume density in the plane of the sky, i.e., 
\begin{equation}
 \Sigma(R) \simeq \sum_{i=3}^n\,a_i\, S_i(R),
  \label{eq:master3}
\end{equation}
\begin{equation}
 S_i(R) = \int_0^{1}\,\alpha^i\,\frac{\xi_\Sigma(\alpha\epsilon_{max},R)}{\sqrt{\epsilon_{max}}}\,d\alpha,
  \label{eq:master4}
\end{equation}
where $\Sigma(R)$ and $\xi_\Sigma(\epsilon_i,R)$ stand the 2D projection (i.e., the Abel transform) of $\rho(r)$ and $\xi(\epsilon_i,r)$, respectively.  $R$  represents for the radial coordinate in the plane of the sky projection, as in Sect.~\ref{sec:observations}. 
%

%
\subsection{Actual algorithm to infer $f(\epsilon)$ from $\Sigma(R)$}\label{sec:the_actual_algorithm}

Except for an arbitrary scaling parameterized by $\epsilon_{max}$, Eqs.~(\ref{eq:polydef}) and (\ref{eq:master3}) provide a method to infer the DF $f(\epsilon)$ needed for a galaxy of observed mass surface density $\Sigma(R)$ to live in a given gravitational potential. A fitting algorithm using  Eq.~(\ref{eq:master3}) provides the coefficients $a_i$ determining $f(\epsilon)$ through Eq.~(\ref{eq:polydef}). The characteristic densities in Eq.~(\ref{eq:master4}) have to be computed numerically starting from the potential in a chain requiring  at least two integrations: the Abel transform that projects the volume densities on the plane of the sky and the integral over all energies expressed by Eq.~(\ref{eq:master4}). We compute the Abel transform using the direct method implemented in the {\tt PyAbel Python} package \citep{2019RScI...90f5115H}. Then the 2nd integration is carried out using the {\tt Simpson}'s rule from {\tt Scipy} \citep{2020SciPy-NMeth}. 
The free parameters retrieved from fitting $\Sigma(R)$ are the amplitudes $a_i$ together with the global radial scaling  factor setting the width of the potential $r_{sp}$ (see Appendix~\ref{sec:plummer_pot}), the latter making the fit non-linear. The fits were carried out using a Bayesian approach. After the initialization using an unconstrained least squares routine \citep[{\tt least\_squares} from {\tt scipy};][]{2020SciPy-NMeth}, the posterior is explored using the   ensemble sampler for Markov Chain Monte Carlo (MCMC) {\tt emcee} \citep[][]{2013PASP..125..306F}. 
Several trial and error tests led us to set the order of the polynomial to 10, a value that provides enough flexibility to reproduce the inner plateau of the observed $\Sigma(R)$ (Fig.~\ref{fig:read_richstein5}). The priors in the Bayesian analysis are uninformative for $r_{sp}$ and $a_i$ ($0 < r_{sp}/b \le 10^3$ and $10^{-2} < |a_i| < 10^2$ relative to the values from the least squares best fit). We also ask the outermost slope of the fitted  $\log\Sigma(R)$ to be less than -2, thus preventing $\Sigma(R)$ to have infinite mass outside the observed radii. In addition, we force $f\ge 0$ so that potential and observation are physically consistent.  All in all, the fits have 9 free parameters (eight $a_i$ plus $r_{sp}$) which is much smaller than the 207 observed points in Fig.~\ref{fig:read_richstein5}.
The posterior was explored with 32 walkers and 6000 samples -- none of the results reported below depend on these exact values. 
%

The algorithm passed a number of sanity checks with systems were the DM distribution is known, namely,  globular clusters and simulated dwarf galaxies. In addition, back-of-the envelope estimates assure the stars in UDFs to be collisionless, as required by EIM.

%
\section{Results}\label{sec:results}
\begin{figure*}[ht!] 
\centering
\includegraphics[width=0.8\linewidth]{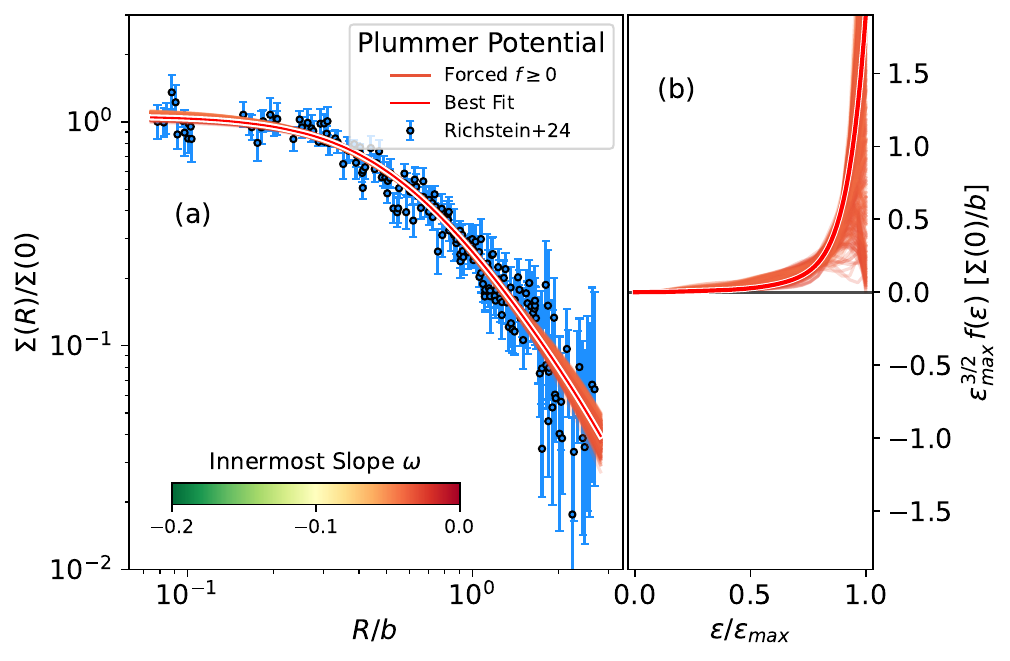}
\caption{
  (a)  Fits to the data in Fig.~\ref{fig:read_richstein5} using $f(\epsilon)$  as free parameter and assuming  the galaxies to reside in a Schuster-Plummer gravitational potential. The best fit is shown as a solid red line. The fits forced to have $f\geq 0$ are shown as colored solid lines, where the color code represents the innermost slope ($d\log\Sigma/d\log R$ when $R\to 0$) as indicated in the color bar.
(b)  $f(\epsilon)$ corresponding to the fits in (a) and using the same color code. Note that the best unconstrained fit yields $f\geq 0$ everywhere.  
}
\label{fig:df3_run_plot_1}
\end{figure*}
\begin{figure*}[ht!] 
\centering
\includegraphics[width=0.8\linewidth]{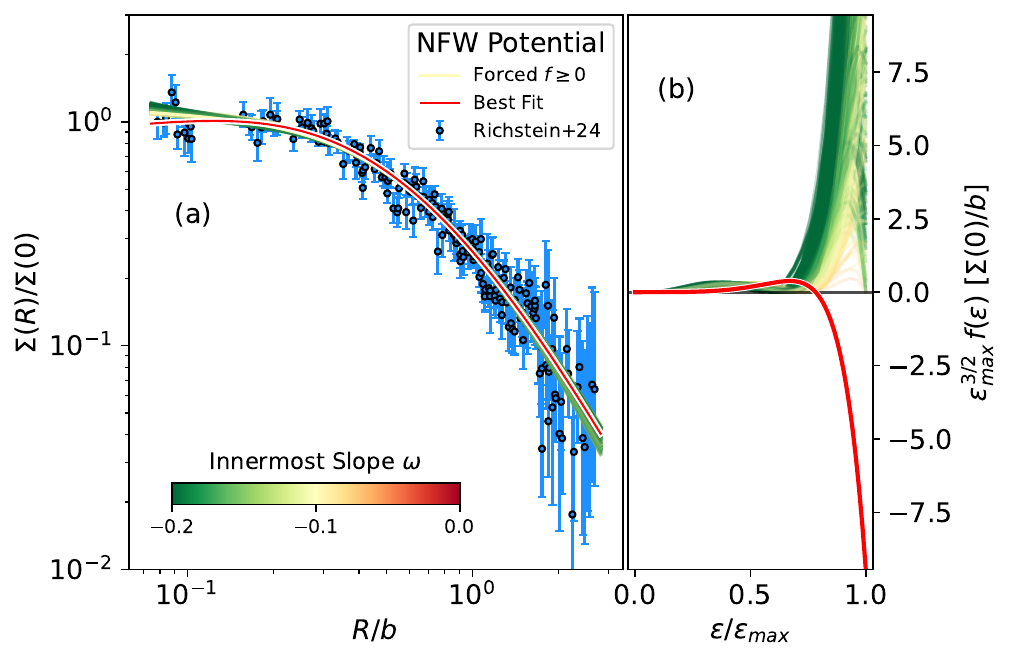}
\caption{
Same as Fig.~\ref{fig:df3_run_plot_1} but assuming the galaxies to reside in a NFW potential. The color code is the same as that employed in Fig.~\ref{fig:df3_run_plot_1}. Note that the best fit requires an unphysical $f < 0$ (the red solid line in panel b), and that the fits forced to have $f \geq 0$, contrarily to the observation,  present quite negative inner slopes  $\omega$ (the coloring is green-yellow rather than orange-red). 
}
\label{fig:df3_run_plot_2}
\end{figure*}
The DF fitting algorithm  in Sect.~\ref{sec:derivation} was applied to the stellar surface density data of \citet{2024ApJ...967...72R} re-scaled as in Fig.~\ref{fig:read_richstein5}. Thus, we consider the observed profile to represent a spherically symmetric galaxy and assume its velocities to be isotropic, assumptions critically inspected in Sect.~\ref{sec:conclusions}.
The results considering a Schuster-Plummer potential (core) and a NFW potential (cusp) are shown in Figs.~\ref{fig:df3_run_plot_1} and \ref{fig:df3_run_plot_2}, respectively. There are two clear differences between them: (1) the best-fitting NFW potential needs and unphysical $f < 0$  (Fig.~\ref{fig:df3_run_plot_2}b, the red solid line) which is not required in the case of a Schuster-Plummer potential (Fig.~\ref{fig:df3_run_plot_1}b). (2) The innermost slope obtained when the fits are forced to have physically sensible $f\geq  0$ are distinctly negative for the NFW potential and near zero for the Schuster-Plummer potential (cf.~the coloring of the thin lines in Figs.~\ref{fig:df3_run_plot_1} and \ref{fig:df3_run_plot_2}). 
\begin{figure*}
\centering
\includegraphics[width=0.8\linewidth]{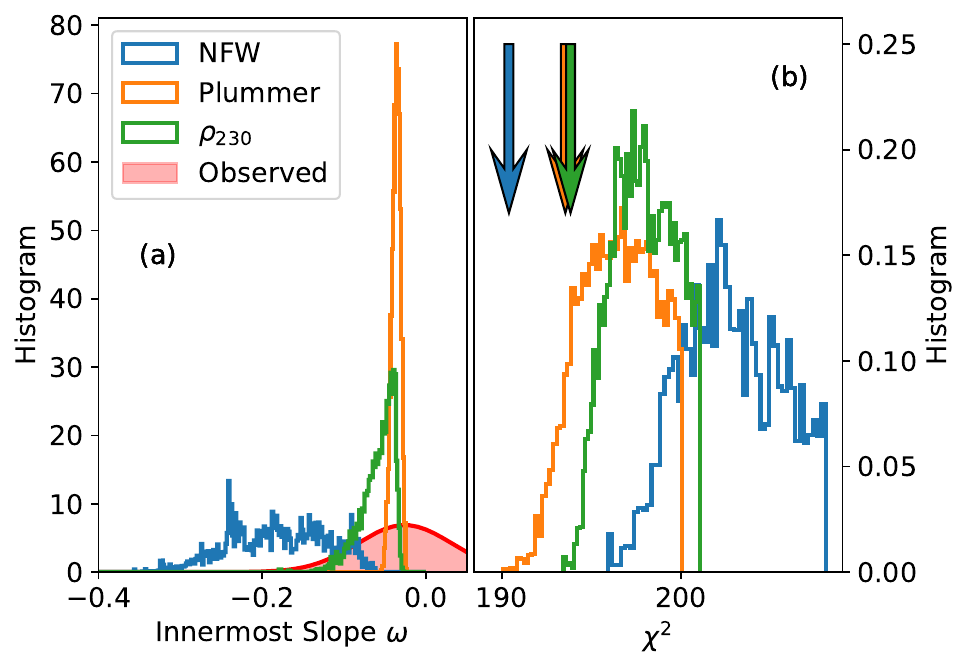} 
\caption{Summary plot used to estimate confidence limits. (a) Histograms with the innermost slopes for the fits with the three gravitational potentials explored in the work (see the inset). The red histogram represents the observed value (Eq.~[\ref{eq:innermostslope}]). (b) Distribution of $\chi^2$ of the fits for the three potentials when forcing $f\geq 0$. The arrows represent the $\chi^2$  of the best fit obtained with unconstrained $f(\epsilon)$. The color code is the same in (a) and (b).}
\label{fig:df3_run_plotf_a}
\end{figure*}
Two other more subtle differences are brought out  in Fig.~\ref{fig:df3_run_plotf_a}:
(3) the $f \geq 0$ Schuster-Plummer potential fits are significantly better than the corresponding NFW potential fits (cf.~their $\chi^2$ in  Fig.~\ref{fig:df3_run_plotf_a}b) and (4) the distribution of innermost slopes of the NFW  potential fits are in tension with the observed innermost slope (Eq.~[\ref{eq:innermostslope}]); cf. the red and the blue histograms in Fig.~\ref{fig:df3_run_plotf_a}a. This tension goes away in the case of a Schuster-Plummer potential (the orange histogram in Fig.~\ref{fig:df3_run_plotf_a}a).
We also analyze the observed stellar surface density  assuming the gravitational potential stemming from a density profile $\rho_{230}\propto 1/(1+r^2)^{3/2}$,  which is similar to Schuster-Plummer in the center and to NFW in the outskirts (see the pink dashed line in Fig.~\ref{fig:df3_run_plotd}). It is also consistent with the observations, very much like the Schuster-Plummer potential (Fig.~\ref{fig:df3_run_plotf_a}).

The distributions in Fig.~\ref{fig:df3_run_plotf_a} are used to work out confidence levels discarding the observed \ufd s to reside in NFW potential under the assumption of spherical symmetry and velocity isotropy. The  $\chi^2$ of the best fitting function is $4.5 \sigma$ off the mean of the $\chi^2$ corresponding to the $f\geq 0$ NFW potential fits (cf.~blue arrow and histogram in Fig.~\ref{fig:df3_run_plotf_a}b). (Here and throughout, $\sigma$ represents the standard deviation of the named distribution.) The best NFW fit has been constructed to have the lowest $\chi^2$, but only in the case of the NFW potential the best-fit does not overlap with the distribution of  $\chi^2$ for $f\geq 0$ (Fig.~\ref{fig:df3_run_plotf_a}b).
%
Assuming that a $2\sigma$ decrease  of the best-fit $\chi^2$ would still be consistent with the $f\geq 0$ $\chi^2$ distribution, we can set the confidence level as follows. Since  NFW best-fit $\chi^2$ is 4.5$\sigma$ off, it wold require an extra 2.5$\sigma$ fluctuation.
Being conservative and assuming a Gaussian tail for the $\chi^2$ distribution, a 2.5$\sigma$ fluctuation has a probability of 0.6\,\%, therefore, we can discard the observed \ufd s to reside in a NFW potential with a 99.4\,\% confidence. The same argumentation applied to the Schuster-Plummer and the  $\rho_{230}$ potentials yields full consistency of the best fits with $f\geq 0$ fits.
A similar exercise can be carried out with the distribution of innermost slopes $\omega$ represented in Fig.~\ref{fig:df3_run_plotf_a}a. The  $f\geq 0$ NFW potential fits yield a distribution with mean and standard deviation of $-0.18\pm 0.06$, therefore, its mean  is $2.7\sigma$ away from the observed value (Eq.~[\ref{eq:innermostslope}]). It is off by only $0.18\sigma$ and $0.58\sigma$ in the case of Schuster-Plummer and $\rho_{230}$, respectively (Fig.~\ref{fig:df3_run_plotf_a}a). Assuming a Gaussian tail for the distribution of errors, the probability for fluctuation $>2.7\sigma$ is 0.004 which would discard the existence of a NFW potential with 99.6\,\%\ confidence. A more conservative approach of setting a confidence level is computing  the probability that the innermost slopes are consistent with the observed slope within $1\sigma$, i.e., $\omega \geq -0.084$. It is only 3\% for the NFW potential whereas it is 100\,\% for the Schuster-Plummer potential and 87\,\% for the $\rho_{230}$ potential. The 3\% value discards  the observed galaxies to reside in a NFW potential with 97\,\% confidence.

\begin{figure}
\centering
\includegraphics[width=\linewidth]{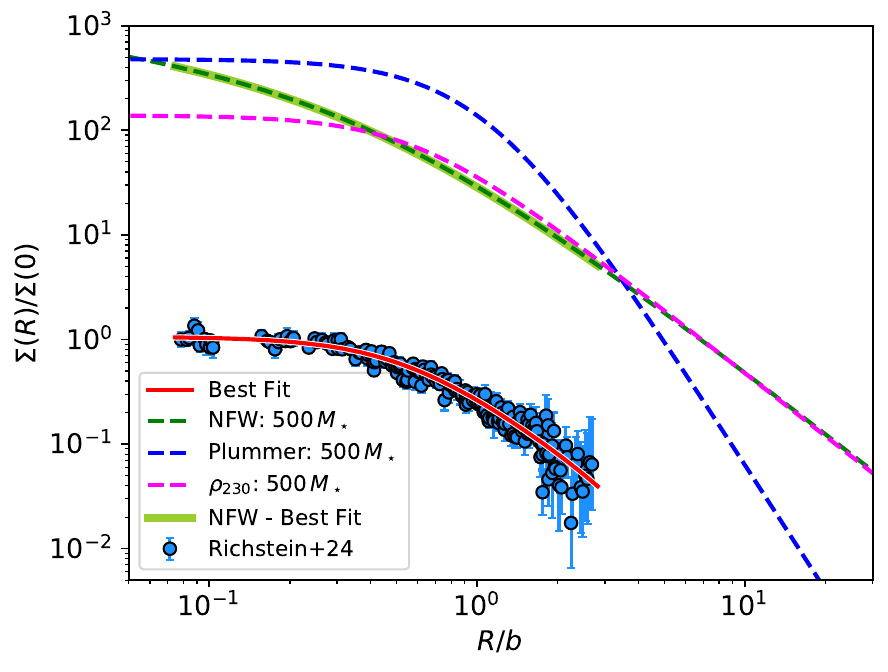} 
\caption{Observed stellar surface density (the symbols) and its best fit assuming a Schuster-Plummer potential (the red solid line). The plot shows the mass surface density that gives rise to the best fitting Schuster-Plummer potential (the blue dashed line), the NFW potential (the green dashed line), and the $\rho_{230}$ potential  (the magenta dashed line). The vertical scaling was chosen to be representative of the UFDs, so that the mass creating the potential is $500$ times the observed stellar mass. The contribution from the stars to the overall potential has been subtracted out in the solid yellow-green line. The result is virtually indistinguishable from the original profile (the green dashed line), showing the selft-gravity of the stars to be negligible.
}
\label{fig:df3_run_plotd}
\end{figure}
Figure~\ref{fig:df3_run_plotd} shows the correspondence between the stellar surface density profile and the mass surface densities producing the best-fitting potentials. The global scaling factor remains unconstrained in our procedure (Sect.~\ref{sec:the_actual_algorithm}), and has been set so that the mass giving rise to the potential is $500$ times $M_\star$ (representative of the observations; see Sect.~\ref{sec:observations}). Interestingly, the core radius $R_c$ of the stars and the potentials are very similar. Defining as core radius when the surface density is half the maximum value, $\Sigma(R_c)=\Sigma(0)/2$,
\def\caca{$\rho_{230} $} 
\begin{equation}
  \log\left[R_{c}^{\rm Potential}/R_{c}^{\star}\right]=
  \cases{~~0.06\pm 0.08 & Plummer, \cr -0.01 \pm 0.13 & {\caca},}
  \label{eq:tricks}
\end{equation}
where $R_{c}^{\star}$ and $R_c^{\rm Potential}$ stand for the core radius of the stars and the potential, respectively.  
The error bars comes from the scatter of the MCMC sampling of the posterior.

%
\section{Discussion and conclusions}\label{sec:conclusions}

The main finding of this study is that six small \ufd\ galaxies do not reside in NFW potentials, a conclusion supported with a confidence level  $\geq 97\,\%$ (Sect.~\ref{sec:results}). The stellar mass of these systems is as low as $10^3$\,--\,$10^4\,M_\odot$, for which baryon feedback should be unable to modify the shape of the CDM potential (Sect.~\ref{sec:intro}), usually represented by a NFW potential. Simultaneously, the observed \ufd s are consistent with potentials with an inner core as predicted by many alternatives to CDM.  Taken at face value, this result point to the DM deviating from collisionless CDM. However, open to scrutiny, the used analysis hinges on several assumptions that could blur such seemingly clearcut evidence. Below we examine the assumptions to conclude that  a deviation from the CDM paradigm still seems to be the main interpretation of the observed stellar cores.

(1) The main observational constraint disfavoring NFW potentials is the innermost slope of $\Sigma(R), \omega,$ being $\simeq 0$ (Eq.~[\ref{eq:innermostslope}]). The constraint is fairly robust largely independent of the method to compute azimuthal average profiles (Sect.~\ref{sec:observations}). The stacking to produce the reference observed profile (Fig.~\ref{fig:read_richstein5}) does not influence it either. We tried several alternatives, and they all render cored stellar surface density profiles which, subject to our analysis, are incompatible with NFW potentials.  

(2) The EIM adopted in the work assumes the velocities to be isotropic (Eq.~[\ref{eq:leading}]). However restrictive, this assumption does not seem to produce the tension between stellar cores and NFW potentials because the tension remains even when the assumption is dropped. The theorem posed by \citet{2006ApJ...642..752A} relates the innermost slope of stars with the velocity anisotropy parameter, discarding all radially biased velocity distributions provided $\omega\simeq 0$.  The Osipkov-Merrit model describes a case in between isotropic and radial,   with isotropic orbits at the center that progressively become radial in the outskirts, and it is inconsistent with NFW potentials too \citep{2023ApJ...954..153S}. Circular orbits offer a chance to reconcile NFW potentials with stellar cores \citep[e.g.,][]{2023ApJ...954..153S}, however, they are expected to be uncommon among the smallest galaxies. Both, the hierarchical growth of galaxies by accretion and the outflows driven by central starbursts cause radial rather than tangential motions. This is indeed found in cosmological numerical simulations of dwarf galaxy formation, which produce Osipkov-Merrit like velocity anisotropy  \citep[e.g.,][]{2017ApJ...835..193E,2023MNRAS.525.3516O} or quasi isotropic orbits \citep[][]{2017MNRAS.472.4786G}. This trend is also found in dwarf spheroidal galaxies with observed kinematics \citep[e.g.,][]{2020A&A...633A..36M,2022A&A...659A.119K}.  More complex anisotropies still need to be fully discarded \citep[e.g.,][]{2017ApJ...838..123S}.

%
(3) The fact that the analyzed galaxies are satellites rather than centrals may have changed the shape of their DM potential through tidal forces with the central galaxy and other satellites. Two arguments seem to minimize the influence of these interactions. First, elaborate CDM-only cosmological numerical simulations show the DM haloes to maintain their identity and the same shape along 30 orders of magnitude in mass \citep{2018MNRAS.474.3043V,2020Natur.585...39W}. Since small haloes are satellites, this simulation shows that the tidal forces arising from the main source of gravitation (i.e., from DM) do not change the shape to be expected in CDM satellite haloes. Second, the history of tidal disruptions suffered by different satellites is different so that different satellites should present different profiles if the shape were set by the tidal influence, and this is not the case with the  stellar distribution in \ufd s (Fig.~\ref{fig:read_richstein5}).  

(4) Spherical symmetry of both DM and stars is assumed  in the EIM analysis. In principle, this is not consistent with the fact that the observed axial ratio of the UFDs often differs from 1 (Sect.~\ref{sec:observations}). However, we note that all observed stellar density profiles collapse to the same profile within errors (Fig.~\ref{fig:read_richstein5}). Since this profile is independent of that axial ratio, it is the one to be expected from a purely spherical stellar system, for which EIM applies consistently. Moreover, one of the targets (\sagii) is round within errors \citep{2024ApJ...967...72R}. We analyze it individually with a result consistent with the whole set: NFW potential fits are significantly worst than Schuster-Plummer  and the required innermost slope $\omega$ differs from zero when forcing $f \geq 0$. 
In addition,  \citet{sanchez_almeida_carlstein} show how the incompatibility between NFW potentials and stellar cores also remains for axi-symmetric systems suggesting that it is more fundamental than, and not attached to, the spherical symmetry assumption. The idea that the incompatibility is not due to the spherical symmetry assumption is also advanced in \citet[][]{2009ApJ...701.1500A}.

(5) As soon as they possess cores,  the actual details of the potential are not important to grant  agreement with stellar cores. This conclusion is attested by the agreement of the observations with both Schuster-Plummer and $\rho_{230}$ potentials (Figs.~\ref{fig:df3_run_plotf_a} and \ref{fig:df3_run_plotd}), and is also by the battery of tests carried out for other potentials by \citet{2023ApJ...954..153S}. The use of NFW to represent CDM  also seems to be unessential since the Einasto profiles, which are also a good representation of the CDM halos, are also incompatible with stellar cores  \citep{2024RNAAS...8..167S}. The Einasto profiles do not diverge at the center indicating that the incompatibility is not artificially set by the mathematical singularity of the NFW profiles when $r\to 0$.

  (6) The hypothesis that stellar feedback is unable to modify the gravitational potential of the observed UFDs is backed up by cosmological numerical simulations (Sect.~\ref{sec:intro}), therefore, it depends on the assumed sub-grid physics for the feedback. Increasing the effectiveness of the feedback may reduce the stellar mass threshold able to influence, but not much. The order of magnitude estimate by \citet[][Fig.~2]{2012ApJ...759L..42P} shows that the threshold is set by energy conservation --  there is not enough energy in supernovas to turn NFW potentials into core potentials when $M_\star \ll 10^6 M_\odot$. The UFDs that we analyze are more than two order of magnitude less massive than this limit and so the ineffectiveness of stellar feedback seems to be granted.

(7) The self gravity of the stars is not considered in our analysis, implicitly assuming the stellar mass to be negligible compared to the DM mass. This is a good approximation for most UFDs (Sect.~\ref{sec:observations}), and since the observed surface densities are the same for all (Fig.~\ref{fig:read_richstein5}), this assumption seems to be safe. 

(8) The fact that \sagii\ may be an extended GC (Sect.~\ref{sec:observations}) is not relevant for the analysis. The whole procedure was repeated withdrawing \sagii\  without any difference that could modify the above conclusions.

To sum up, this work shows that six \ufd\ galaxies reside in cored gravitational potentials. Since stellar feedback should be inoperative in their stellar mass regime (HUG),  the best explanation seems to be that the DM deviates from the nature assumed in the standard  $\Lambda$CDM cosmological model.  The standard model provides an extremely good approximation to reality but is likely not the last theory \cite[][]{2021arXiv210602672P}. Studying the kind of galaxies analyzed here may provide a gateway to go beyond.

\begin{acknowledgments}
This work is indebted to the many colleagues whose comments and criticisms help us to sharpen our arguments, in particular, Diego Blas,  Kfir Blum, Andrea Caputo, Christopher Garling, Nitya Kallivayalil, Jorge Mart\'\i n Camalich,  Justin Read, Hannah Richstein, Jack Warfield, and an anonymous referee. Richstein and Kallivayalil kindly provided the exceptional dataset used in the work.
JSA acknowledges financial support from the Spanish Ministry of Science and Innovation, project PID2022-136598NB-C31 (ESTALLIDOS). 
IT acknowledges support from the ACIISI, Consejer\'{i}a de Econom\'{i}a, Conocimiento y Empleo del Gobierno de Canarias and the European Regional Development Fund (ERDF) under grant with reference PROID2021010044 and from the State Research Agency (AEI-MCINN) of the Spanish Ministry of Science and Innovation under the grant PID2022-140869NB-I00 and IAC project P/302302, financed by the Ministry of Science and Innovation, through the State Budget and by the Canary Islands Department of Economy, Knowledge and Employment, through the Regional Budget of the Autonomous Community.
The HST data were observed as part of Treasury Program GO-14734 (PI Kallivayalil). Support for this program was provided by NASA through grants from the Space Telescope Science Institute, which is operated by the Association of Universities for Research in Astronomy, Incorporated, under NASA contract NAS5- 26555.
\end{acknowledgments}

%

\vspace{5mm}
\facilities{HST (WFC\,--\,WFC3)}


\software{{\tt PyAbel} \citep{2019RScI...90f5115H}, {\tt Scipy} \citep{2020SciPy-NMeth},   {\tt emcee} \citep[][]{2013PASP..125..306F}. 
          }






\appendix
\section{Characteristic densities in the case of a Schuster-Plummer potential}\label{sec:plummer_pot}
Consider the gravitational potential generated by a mass distribution following a Schuster-Plummer density profile, i.e.,
\begin{equation}
  \rho_{\rm SP}(r) = \frac{\rho_{sp}}{\left[1+(r/r_{sp})^2\right]^{5/2}},
\end{equation}
where $\rho_{sp}$ and $r_{sp}$ are the central density and the characteristic radial scale, respectively. The gravitational  potential produced by this {\em cored} density profile is  (e.g., \citeauthor{2023ApJ...954..153S}~\citeyear{2023ApJ...954..153S}, Eq. [A14]),
\begin{equation}
  \Psi_{\rm SP}(r) = \frac{\epsilon_{max}}{\left[1+(r/r_{sp})^2\right]^{1/2}},
\end{equation}
with $\epsilon_{max}=4\pi G \rho_{sp}r_{sp}^2/3$. Equation~(\ref{eq:master_mind}) renders,
\begin{equation}
  \xi(\epsilon,r)= \sqrt{\epsilon_{max}}\, h(\epsilon/\epsilon_{max},r/r_{sp}),
\end{equation}
\begin{equation}
  h(\alpha,\beta) = 4 \pi \sqrt{2}\,\sqrt{\left[\left(1+\beta^2\right)^{-1/2}-\alpha\right]}\,\Pi\left(\beta-\beta_X\right),
  \label{eq:master_sp}
\end{equation}
with $\beta_X = \sqrt{1-\alpha^2}/\alpha$.

\end{document}